\documentstyle[floats,prd,aps,epsfig]{revtex}
\newcommand{\bce}{\begin{center}} \newcommand{\ece}{\end{center}}
\newcommand{\beq}{\begin{equation}} \newcommand{\eeq}{\end{equation}}
\newcommand{\beqy}{\begin{eqnarray}}
\newcommand{\eeqy}{\end{eqnarray}}

\input epsf
\advance\voffset by 0.5in
\newcommand{\ome}{\omega_B}

\advance\voffset by -0.5in
\begin{document}
\title{Surface structure of quark stars with magnetic fields}
\author{Prashanth Jaikumar\footnote{jaikumar@phy.anl.gov} 
%and Rachid Ouyed$^2$
}

\address{Physics Division, Argonne National Laboratory, 
Argonne IL 60439, USA.\\
         %$^2$Department of Physics and Astronomy, University of Calgary,
% Calgary, AB T2N 1N4 Canada\\
}

\maketitle
\begin{abstract}
We investigate the impact of magnetic fields on the electron
distribution in the electrosphere of quark stars. For moderately
strong magnetic fields $B\sim 10^{13}$G, quantization effects are
generally weak due to the large number density of electrons at
surface, but can nevertheless affect the spectral features of
quark stars. We outline the main observational characteristics of
quark stars as determined by their surface emission, and briefly
discuss their formation in explosive events termed Quark-Novae, which
may be connected to the $r$-process.  \\
\bigskip


\end{abstract}
%------------------------------END OF TITLE AND ABSTRACT------------

%-----------------------------BEGIN INTRODUCTION-------------------

\section{Introduction}
        \label{sec_intro}
There is renewed interest in the theory and observation of strange
quark stars, which are believed to contain, or be entirely composed
of, deconfined quark matter~\cite{JSB}. An observational confirmation
of their existence would be conclusive evidence of quark deconfinement
at large baryon densities, an expected feature of Quantum
Chromodynamics (QCD). Furthermore, discovery of a stable bare quark star
affirms the Bodmer-Terazawa-Witten conjecture~\cite{BTW}, viz., that
at high enough density, strange quark matter, composed of up, down and
strange quarks, is absolutely stable with respect to nuclear
matter. This intriguing hypothesis is over three decades old, and bare
quark stars are but one possible realization put forward in the
intervening years. Broadly speaking, the field of quark star research
follows two intertwined paths: (1) efforts to describe the global
(mass and radius) and surface structure of such stars from
microscopics and (2) determination of observational features that
distinguish them from their close cousins, neutron stars. It is
possible that a few of the approximately 1500 neutron stars detected
thus far are really quark stars~\cite{Xu1}.

\vskip 0.2cm
 
A positive identification of a quark star would likely require a
complement of consistent signals. For example, physical properties
such as maximum mass, radius, minimum spin-periods and neutrino/photon
cooling rates (and hence temperature) all depend on the (unique)
equation of state for dense quark matter. As the latter is uncertain,
there is a range of allowed values for the above quantities that makes
a distinction based on current observational data difficult. We
emphasize that a bare strange star is self-bound~\cite{BTW,Chin79}
while a neutron star requires gravitational forces for its
binding. This implies a very different equation of state for the two
classes of stars; however, in the range of masses
($1.35<M/M_{\odot}<2.4$) observed to date, the calculated radii are
similar ($R\sim 10$ to 15 km). Perturbative corrections to the quark
matter equation of state and color superconductivity effects further
complicate the issue~\cite{ARB}.~\footnote{We will not elaborate here
on the possibility of having superconducting phases of quarks. We
refer the interested reader to Refs.~\cite{ABR,IS,Ferrer,Blaschke,ALP}
for a consideration of these phases, and the consequences of their
existence inside compact stars in the present context.}

\vskip 0.2cm 

It was first pointed out by Alcock et al.~\cite{AFO} that the bare
quark star surface would be characterized by an abrupt termination of
the quark density (density gradient $\sim 10^{26}$g/cm$^4$) and a less
rapid fall-off of a charge neutralizing skin of electrons termed the
electrosphere (density gradient $\sim 10^{18}$g/cm$^4$). The latter
extends up to 1000 Fermis above the bare quark star surface. The
electric field which binds these electrons to the surface is large
enough to pull electron-positron pairs out of the QED vacuum,
prompting copious pair-emission which is directed outward by the
radiation output by the star~\cite{AMU}. This mechanism of emission
from the bare surface of a strange quark star, due to both photons and
$e^+e^-$ pairs, can produce luminosities well above the Eddington
limit ($\sim 10^{38}~{\rm erg~sec}^{-1}$) up to a decade after the
quark star's birth~\cite{Page02}. The spectrum of emitted photons is
significantly harder than that from a normal cooling neutron star ($30
<E/{\rm keV} < 500$ instead of $0.1 < E/{\rm keV} <2.5$). After the
star has cooled to temperatures less than $T< 6\times 10^8$K, other
microscopic processes of photon emission can dominate the luminosity
of the star and yield a non-thermal spectrum~\cite{JPPG}. This
distinctive spectrum and temperature evolution, if observed, would
constitute an unmistakable signature of a strange quark star, but
would require a fortuitous detection before the star cools below the
observational threshold of present-generation space-based laboratories
such as INTEGRAL~\cite{Int}.

\vskip 0.2cm

Quark stars may be much more elusive if they have a suspended crust
of normal matter up to the neutron drip density, that is separated from
the quark surface by a microscopic gap but supported by the large
electric fields at surface~\cite{AFO,Madsen}. Alternatively, quark stars
may admit a mixed phase crust several meters in thickness and
extending up to its surface, if surface and Coulomb costs (determined
at the microscopic level by the current quark mass of the strange
quark) is small enough~\cite{JRS,ARRS}. This would obviate the need for
large electric fields at surface, smoothen the density gradient at
surface, and render the previously proffered signals
inapplicable. These uncertainties are not expected to be resolved
quickly since they involve quantifying the parameters of QCD to better
than 5\% in a distinctly non-perturbative regime. (NB. The density that
characterizes quark matter in these objects corresponds to a quark
chemical potential $\mu_q\sim 500$ MeV which does not greatly exceed
$\Lambda_{\rm QCD}\sim 200$ MeV.)

\vskip 0.2cm

However, we can certainly aim to quantitatively improve our current
understanding of the surface structure under the working assumption
that the star has a bare quark surface with an electrosphere, since
electromagnetic effects dominate over QCD effects. A clear and
necessary improvement is the introduction of surface magnetic fields.
Neutron stars are born with intense magnetic fields of order $10^{12}$
Gauss or higher, which may decay by over 5 orders of magnitude in the
star's lifetime. If a neutron star undergoes a phase conversion to a
bare quark star, the surface magnetic fields are expected to remain
intact in geometry, unless interior magnetic fields are expelled by a
superconducting state that displays the Meissner effect, in which case
rapid magnetic reconnections occur, diminishing the field at surface
to almost zero~\cite{OYD1}. In this work, we will superimpose a static
magnetic field $B$ on the electrosphere of a quark star, and calculate
the profile for the electrosphere as a first step toward
characterizing the spectrum of a magnetized quark star. We expect
considerable differences from the $B=0$ case since magnetized
atmospheres of neutron stars are already known to be very different
from non-magnetized ones~\cite{Pons}. Further, the inclusion of
magnetic fields is a step toward a more realistic treatment of the
quark star spectrum.

\vskip 0.2cm

In the next section, we reintroduce the established picture of the
electrosphere for the purpose of a self-contained presentation, and
mention the modifications due to the strong magnetic field. In section
III, we quantify the electrostatic potential and electron number
density in the magnetized electrosphere, and state qualitative impacts
on the electrospheric photon emission. We conclude in section IV with
a general discussion on the formation of Quark stars, with particular
attention to the attractive Quark-Nova scenario which may be connected to
$r$-process nucleosynthesis.

%-----------------------END OF INTRODUCTION--------------------%
%-----------------------BEGIN SECTION II-----------------------%

\section{Degenerate electron gas in a strong magnetic field}
         \label{sec:Belectrosphere}
Recently, a few authors~\cite{UCH} pointed out that the deficit of
(massive) strange quarks due to surface effects on the star can lead
to a thin charged skin a few Fermis thick, which can lead to a jump in
the electric field at surface. The local charge neutrality condition
is $2/3n_u-1/3n_d-1/3n_s-n_e=0$, with the appropriate low temperature
($T\ll \mu_i$) expressions in the bulk:

\beq
n_i=\frac{p_{F,i}^3}{\pi^2}=\frac{(\mu_i^2-m_i^2)^{3/2}}{\pi^2};\quad m_e\ll m_{u,d}\ll m_s<\mu_q ~\forall q\quad .
\eeq

Imposing the conditions of $\beta$-equilibrium and baryon number
conservation shows that $\mu_e\simeq m_s^2/4\mu$ to better than
1\%. At surface, $s$-quarks are depleted due to surface tension
effects by an amount~\cite{mad}

\beq
\delta n_s = -\frac{3}{4\pi}p_{F,s}^2\psi(\lambda_s); ~\psi(\lambda_s) = \left[\frac{1}{2}+\frac{\lambda_s}{\pi}-\frac{(1+\lambda_s^2)}{\pi}{\rm arctan(\lambda_s^{-1})}\right]; ~\lambda_s = \frac{m_s}{p_{F,s}}\,,
\eeq   

leading to a charged layer at the surface of a bare quark star
characterized by a surface charge density $\sigma = -(1/3)e\delta n_s$
and a discontinuous normal electric field with $\Delta E =
4\pi\sigma$. For ``cold'' electrospheres, corresponding to the
temperature bound $T< 5\times 10^{11}$K, the electric field is
directed inward just underneath the surface and outward just above
it. The electrostatic potential and the electron number density are
substantially increased cf. the case when surface effects are
ignored (NB. The corresponding surface depletion of light quarks is
negligible.)~\cite{UCH}.

\vskip 0.2cm 

The large electric field leads to the Schwinger instability of the QED
vacuum~\cite{Schwinger}, resulting in electron-positron pair-emission,
which can be an additional source of photons in the electrosphere,
which normally radiates photons via 2$\rightarrow$ 3 QED processes. A
model cooling calculation of bare strange stars with electrospheric
effects and the inclusion of color superconductivity (which can alter
the specific heat and thermal conductivity of quark matter) has also
been performed~\cite{Page02}. The conclusion is that bare strange
stars will display Super-Eddington photon luminosities at surface
temperatures $T>10^{9}$K, with a spectrum that is hard enough to
distinguish it from thermally radiating neutron stars.  At lower
temperatures $6\times 10^8$K$<T<10^9$K, the bulk of the luminosity
comes from electron-positron pairs which subsequently imprint a wide
annihilation line on the non-thermal spectrum. Thermal emission is
much suppressed owing to plasma frequency effects of quark matter and
the electrosphere~\cite{CH}. Further cooling to temperatures $T<10^8$K
results in a non-thermal spectrum dominated by bremsstrahlung photons
from electron-electron collisions in the electrosphere~\cite{JPPG}.

\vskip 0.2cm

As mentioned in the introduction, a magnetic field is expected to
exist at surface, altering the electrosphere profile and its emission
properties. The electron motion perpendicular to the
magnetic field is quantized into Landau orbitals enumerated by the
quantum number $n_L=0,1,2..$ . For a relativistic fermion of charge
$q$ and mass $m$, such as the quarks and electrons in our case, the energy
in a $\hat{z}$-directed magnetic field $B$ is given by

\beq
\label{benergy}
\epsilon_{n} = \sqrt{m^2c^4+c^2p_z^2+2mc^2\hbar\omega_Bn}\equiv mc^2E;\quad 2n = 2(n_L+\frac{1}{2}+\sigma);\quad \omega_B=\frac{qB}{mc}\,,
\eeq

where $\sigma=\pm 1/2$ is the spin-degeneracy of the fermion. The
$n=0$ state is non-degenerate while $n>0$ states are doubly
degenerate. We express the magnetic field strength in convenient units
'$b$' such that

\beq
b\equiv \frac{\omega_B}{mc^2} = \frac{B}{4.414\times 10^{13}~{\rm G}}\quad{\rm for ~electrons}
\eeq

and take $B=4.414\times 10^{13}$G (corresponding to $b=1$) as an upper
limit on the surface magnetic field of quark stars. Although higher
fields may exist in magnetars and AXPs (anomalous X-ray pulsars),
their interior magnetic fields would likely be large enough to inhibit
the formation of quark matter~\cite{SC1}. For a given magnetic field,
both temperature and matter density determine the filling fraction of
the Landau levels, and hence the degree of quantization. From
Eqn.(\ref{benergy}), the highest Landau level occupied by a fermion
with energy $\epsilon_{n}$ is

\beq n_{L,{\rm max}}: n_{{\rm max}}={\rm Int}(\nu);\quad \nu
=\frac{E^2-1}{2b}\quad .  \eeq

Since the quarks are highly relativistic, $E\gg 1$ and many Landau
levels are occupied for the light quarks, and even more for the
strange quark, so that quantization effects may be ignored.  Further,
when the temperature $T\gg T_B$ where $T_B=\hbar\omega_B/k_B$ ($k_B$
is the Boltzmann constant), thermal effects smear out adjacent Landau
levels, and the magnetic field is effectively non-quantizing,
irrespective of the degeneracy. This effect is more important for
quarks, which have a larger mass, and hence smaller cyclotron
frequency, than electrons. We thus expect no modifications to the
quark distribution due to magnetic fields of typical magnitudes
considered in this work. 

\vskip 0.2cm

For electrons, quantization effects become important as we move
outward from the electrosphere, and more electrons reside in the lower
Landau levels. The number density of electrons in a magnetic field is
given by

\beq
\label{negen}
n_e=\frac{m_e\ome}{h^2}\int_{-\infty}^{\infty}dp_z\sum_{n,\sigma}f(\epsilon_n);\quad f(\epsilon_n)=\frac{1}{1+{\rm e}^{(\epsilon_n-\mu_e)/T}}\quad .
\eeq

For a cold degenerate electron gas which satisfies $T\ll\mu_e$, we may
approximate

\beq
\label{ne}
n_e(\mu_{e0})=\Theta(\mu_{e0}-1)\frac{2m_e^2\ome c}{h^2}\left[\sqrt{\mu_{e0}^2-1}+2\sum_{n=1}^{n_{\rm max}}\sqrt{\mu_{e0}^2-1-2bn}\right]; \quad \mu_{e0}=\frac{\mu_e(T=0)}{m_ec^2}\quad .
\eeq

From Eqn.(\ref{ne}), the density at which the second Landau level
becomes populated is

\beq
n_e^{\ast}=\frac{(m_e\ome)^{3/2}}{\hbar^2\pi^2\sqrt{2}}=1.24\times 10^{30}b^{3/2} ~{\rm cm}^{-3}\quad .
\eeq 

This is already less than the typical electron density at surface in
the $B=0$ case, where $n_e\sim 10^{-4}-10^{-3}/{\rm
fm}^{3}$~\cite{Usov}. Therefore, several Landau levels will be
occupied, and the $B$-field is non-quantizing. In fact, for $\nu\gg 1$
and $T\ll\mu_e$, it follows from Eqn.(\ref{ne}) that~\cite{Shapiro}

\beq
n_e=\frac{2m_e^2\ome c}{h^2}\sqrt{\mu_{e0}^2-1}\left[\frac{2\sqrt{\mu_{e0}^2-1}}{3b}-3\right]\approx \frac{(\mu_e^2-m_e^2)^{3/2}}{3\pi^2}\quad ,
\eeq

which is the usual expression relating the number density and chemical
potential for a relativistic degenerate Fermi gas in the $B=0$
case. For the range of temperatures and densities considered in this
work, $T\ll T_B=\hbar\omega_B/k_B$ holds, so that temperature effects
on electron occupation numbers is negligible. The situation is summarized
in Fig. 1, which shows how the number density of electrons, as
obtained from Eqn.(\ref{negen}) behaves as a function of electron
chemical potential, temperature and magnetic field. The truncation of
the curves at $\nu=0$ for finite $T$ represent the fact that for lower
(negative) chemical potentials, and hence densities, all electrons
reside in the $n=0$ state. We note that only for temperatures $T\geq
T_B$ does the number density deviate significantly from the $T=0$
curve (NB. This is due to the smearing of the Landau levels by
temperature fluctuations.). We therefore continue to use the zero
temperature approximation to the electron density in the rest of the
paper.
\vskip 0.5cm

\begin{figure}[!ht]
\bce
\epsfig{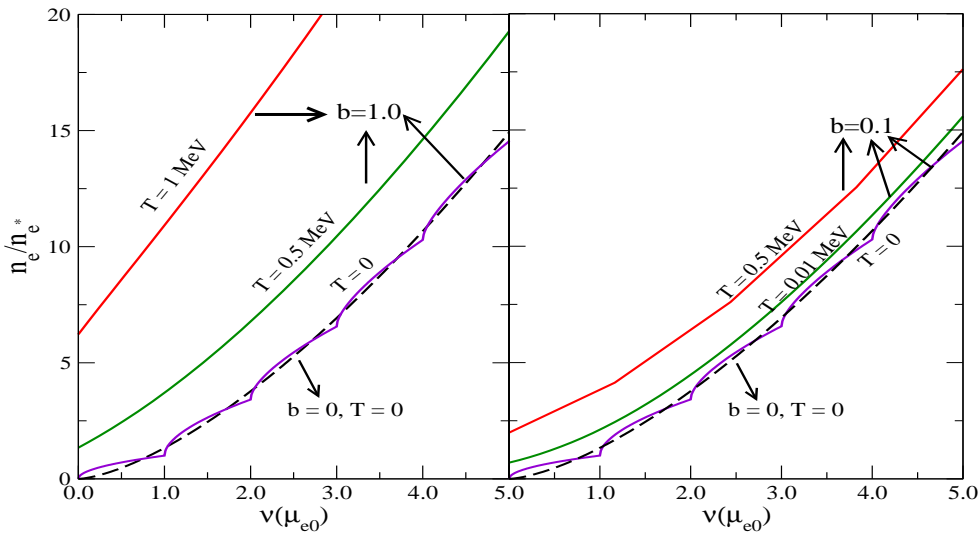}
\ece
\caption{The number density of electrons (scaled to $n_e^{\ast}$) from Eqn.~(\ref{negen}) as a function of the filling factor $\nu$ for various temperatures. The solid curves are for non-zero magnetic field, and display a loss of the Landau level structure with increasing temperature.}
\label{figps1}
\end{figure}

%-----------------------END OF SECTION II---------------------------%occurs in the

%-----------------------BEGIN SECTION III---------------------------%
\section{Electrosphere profile with a magnetic field}
         \label{sec:brem}
We would now like to obtain the electron density profile as a function
of distance in the electrosphere. This will be modified from previous
results~\cite{AFO,CH} which were in the absence of a magnetic
field. The baryon density at the surface of a strange star, where the
quark pressure vanishes, is expected to be in the range $(2-5)\rho_0$,
corresponding to quark chemical potentials in the range 300-400
MeV. At such densities, the strange quark current mass is not
ignorable, with $m_s\sim150$ MeV. Local charge neutrality then implies
an electron chemical potential at surface $\mu_e\approx 15-20$
MeV. This picture of a homogeneous surface as opposed to a crusty one
with quark nuggets is justified by the large value of the surface
tension given our parameter values~\cite{Berger}. The surface
depletion of strange quarks at the surface results in a positively
charged layer, which is two orders of magnitude thinner than the
electrosphere. For cold electrospheres, this charged layer results in
oppositely directed electric fields at the star's surface as well as
an enhanced electron number density $n_e\gg n_{\ast}$.

\vskip 0.2cm

For a general magnetic field with poloidal and toroidal components at
the surface, the electron distribution can, in principle, be obtained
using the formalism of magnetic field density functional theory or by
Green function techniques, both of which are complicated tasks.  As a
first study, we adopt a simple picture of a constant magnetic field
directed normally to the surface and ignore Coulomb interactions and
exchange effects, which are justified approximations given the extreme
degeneracy being explored in the problem. We employ the
self-consistent Thomas-Fermi approximation and present a numerical
solution of the Poisson equation

\beq
\label{eqntfermi}
\nabla^2\phi(\vec{r},\vec{B})=-\rho(\vec{r},\vec{B})=en_e(\vec{r},\vec{B})\,.
\eeq

In the plane parallel approximation $\vec{B}=B\hat{z},~\hat{r}=\hat{z}$, 
and equilibrium requires $\mu_e(n_e(z))-e\phi(z)=0$. For arbitrary magnetic
fields, the equilibrium constraint involves magnetic forces $F_{\rm mag}=
\vec{j}\times\vec{B}\neq 0$ where $\vec{j}$ is the electron current density.
With our approximations, the problem is an electrostatic one, and
the electrostatic potential outside ($z>0$) the star solves

\beqy 
\label{eqntfermi2}
\nabla^2\eta_+&=&\frac{2\alpha
bm_e^2}{\pi}\Theta(\eta_+-1)\left[\sqrt{\eta_+^2-1}+2\sum_{n=1}^{N}\sqrt{\eta_+^2-1-2bn}\right];\quad
N={\rm Int}\left(\frac{\eta_+^2-1}{2b}\right)\quad ,\\ 
\eta_+&=&\frac{e\phi_+}{m_ec^2}; \quad\phi_+=\phi(z>0) \quad .
\eeqy

In the limit $N\gg 1$, when the electrons populate several levels, we recover
the expression for the non-magnetic case

\beq
\nabla^2\phi_+=\frac{4\alpha}{3\pi}e^2\phi_+^3\quad ,
\eeq

while for $N=0$, we obtain 

\beq
\nabla^2\phi_+=\frac{em\omega_Bp_F}{2\pi^2}\quad ,
\eeq

the expression which has been derived previously for the
non-relativistic case~\cite{Fushiki} but applies equally well to the
relativistic case as long as only the 0$^{\rm th}$ level is
populated. It is interesting to note that the above equation leads to
an exponentially decaying electron profile at large distances from the
surface, contrary to the $1/z$ fall-off in the non-magnetic case. As
$\eta_+\rightarrow 1$, we expect interactions and exchange effects to be
important, but since this is only a minute region of the
electrosphere, we do not correct for it.

\begin{figure}[!ht]
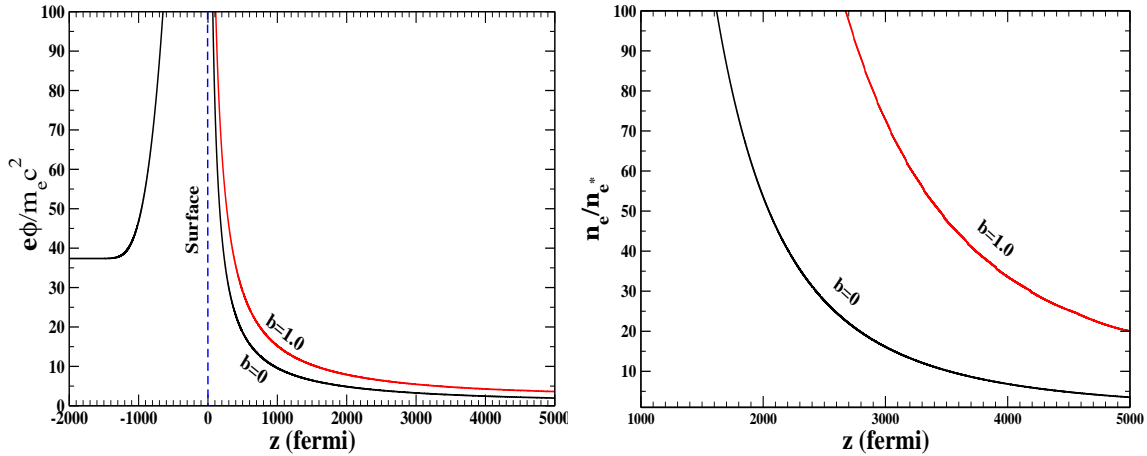

\bce
\leavevmode
\epsfig{file=phi.eps,width=7.5cm,height=6.0cm}
\epsfig{file=tfermi.eps,width=7.5cm,height=6.0cm}
\ece
\caption{Left panel: The electrostatic energy (scaled to $m_ec^2$) at
the edge of the star, from Eqn.~(\ref{eqntfermi}) for the unmagnetized
($b=0$) and magnetized ($b=1.0$) cases. Right panel: The corresponding
electron number density (scaled to $n_e^{\ast}$) at the surface.}
\label{figps2}
\end{figure}

Eq.~(\ref{eqntfermi2}) can be solved numerically once the boundary
conditions are determined. Clearly, $\eta_+\rightarrow 1$ as
$z\rightarrow\infty$ (this corresponds to vanishing electron number
density and is numerically implemented by the shooting method to
obtain the correct asymptote at sufficiently large $z$. This method is
also practically applicable to the case of suspended crusts above a
quark star; see Ref.~\cite{Madsen}.). The second boundary condition is
determined at the surface $z=0$, where
$\phi=\phi_0\Rightarrow\eta=\eta_0$ and $\eta_0$ must be
determined by the discontinuity in the electric field
$\nabla\phi_+-\nabla\phi_-=4\pi\sigma$ with $\phi_-=\phi(z<0)$. The
electrostatic potential $\phi_-$ inside the star is almost identical
to the non-magnetic case since the electron density deep inside the
star, given by $\tilde{\mu_e}=m_s^2/4\mu_q\sim 10~{\rm MeV}\gg
m_ec^2=0.5~{\rm MeV}$ so that for $b\leq 10.0$ fields, quantization
effects are negligible.  In this case, $\phi_-\rightarrow \phi_q$ as
$z\rightarrow -\infty$ where the relation between $\phi_0$ and $\phi_q$
is given by

\beq
4\pi\sigma=e\left(\frac{2\alpha}{3\pi}\right)\left[\phi_0^2+(\phi_q-\phi_0)\left\{(\phi_0+\phi_q)^2+2\phi_q^2\right\}\right]^{1/2}\quad ,
\eeq

wherein the prefactor corrects an error in equation (30) of
Ref.~\cite{UCH}.  For our choice of $\tilde{\mu_e}=11.25$ MeV
(corresponding to $m_s=150$ MeV and $\mu_q=500$ MeV),
$\phi_q=37.4$ MeV, $\phi_0=578$ MeV and hence $\eta_0=341$ MeV. The
boundary conditions having been specified, the results for the
electrostatic potential $\phi_+,\phi_-$ as obtained from
Eq.(\ref{eqntfermi}) are illustrated in Fig.~\ref{figps2}, and
compared to the corresponding expressions for the non-magnetized
electrosphere with the same boundary conditions.

\vskip 0.5cm

We observe that while the electrostatic potential inside the star is
unaffected by magnetic field strengths of $b\sim{\cal O}(1)$, that
outside the star is increased in comparison to the unmagnetized
case. A similar finding for the number density reflects the fact that
the electrons become localized in configuration space, their wave
functions compressed into the Larmor radius, so that the Pauli
repulsion is decreased and electrons can be squeezed together in
greater numbers than in the unmagnetized case. This effect is most
pronounced when the quantization effects are large, as can be seen by
the wiggles that demarcate the filling of the Landau levels (right
panel of Fig.~\ref{figps2}). In both figures above, the curves meet at
the surface ($z=0$) where they satisfy the same boundary condition. At
distances far from the surface, the curves cross each other since the
total number density is the same in both cases. The number density
falls off exponentially for the magnetized case while it falls off as
$1/z$ for the unmagnetized case.

\vskip 0.2cm

In the absence of a thin ($10^{-5}$M$_{\odot}$) crustal layer of
accreted matter above the strange star, the spectral properties of a
strange star are linked to radiative transport in the
electrosphere. This is because radiation from the stellar interior is
cutoff by the high plasma frequency ($\omega_p\sim$ 20 MeV) of dense
quark matter, and Landau-Pomeranchuk-Migdal (LPM) suppression of
bremsstrahlung photons from quark-quark collisions~\cite{CH}. For an
unmagnetized electrosphere, the emission at high luminosities is
dominated by $e^+e^-$ pair emission so long as the temperature is high
enough. At lower luminosities, the emission is dominated by photon
bremsstrahlung from electron-electron collisions in the
electrosphere. The spectrum is non-thermal due to the finite plasma
frequency of the degenerate electron gas and large mean free path for
photons in the electrosphere.

\vskip 0.2cm

The effect of the intense magnetic field on electrospheric emission is
currently being investigated and will be reported elsewhere; here, we
only state some qualitative effects. The plasma frequency plays a
vital role in the emission properties of the electrosphere, since
lower frequency photons are rapidly damped by the complexion of the
refractive index

\beq n_r^2=1-\frac{\omega_p^2}{\omega^2};\quad
\omega_p^2\simeq\frac{e^2}{3\pi^2}\mu_e^2\quad . 
\eeq

This collective effect of the plasma is important at high electron
density (inner regions of the electrosphere) where the Debye length is
smaller than the mean free path. Note that despite the large electron
density, the large degeneracy limits the available scattering states
of electrons, leading to a large mean free path for photons. In the
presence of a magnetic field, when quantization effects become
important, collective effects are determined by the Larmor radius
rather than the Debye length, and the reflectivity for normal
incidence becomes~\cite{Van}

\beq
R=\frac{1}{2}\left[\left|\frac{n_{r,1}-1}{n_{r,1}+1}\right|^2+\left|\frac{n_{r,2}-1}{n_{r,2}+1}\right|^2\right];\quad
n_{r\{1,2\}}^2=1-{\frac{\omega_p^2}{\omega^2}}\frac{\omega}{(\omega\pm\omega_B)}\quad.
\eeq

When $\omega_p\ll\omega\ll\omega_B$, the reflectivity is considerably
diminished from the unmagnetized case, and the electron plasma will
radiate more strongly. Further, the dielectric tensor will modify the
Poisson equation, resulting in the breakdown of the plane parallel
approximation employed here. Thus, our estimate of the profile is
simplistic and must be made self-consistent. The intense magnetic
field will also lead to polarization effects. Taking into account the
thermal conductivity of the electrons in a strong magnetic field, a
non-uniform surface temperature distribution is expected. The
inclusion of toroidal components of the magnetic field can smoothen
out some of these non-uniformities, and has been linked to the largely
uniform blackbody spectrum of some isolated neutron
stars~\cite{GKP1}. It would be interesting to see if the
electrospheric surface of the strange star displays similar spectral
features. In particular, the possibility of having bound positronium
in the magnetized electrosphere should be investigated.

%-----------------------END OF SECTION III-----------------------------%
%-----------------------BEGIN SECTION IV-------------------------------%

\section{Strange stars and Astrophysics: The Quark-Nova and the $r$-process}
         \label{sec:quarks}
Strange quark stars have been linked to some of the most energetic and
violent phenomena in astrophysics, such as Gamma-ray bursts
(GRBs)~\cite{PCH}, type I X-ray superbursts~\cite{Cumming} and
Soft-gamma repeaters (SGRs)~\cite{Usov2} in what is being perceived as
a new paradigm in astrophysics - the connection of high-energy
astrophysics with exotic forms of dense matter. The energy released in
the phase conversion of a hadronic star to a quark star is typically
about $10^{53}$ ergs, similar in magnitude to the energy of observed
gamma-ray bursts, and accretion and flash heating on a strange star
surface is a viable model for SGR flares and superbursts. Further, the
emission spectra of the long bursts can also be explained by invoking
the presence of strange quark matter and its subsequent transition to
a color-superconducting state~\cite{ORV}. All of these scenarios are
predicated upon the existence of absolutely stable three-flavor quark
matter at high density.

\vskip 0.2cm

Here, we discuss the formation of a quark star in a Quark-Nova, a
scenario in which nuclear matter at high density in the core of a
neutron star undergoes a phase transition to two flavor (up and down)
quark matter, which then undergoes equilibrating weak reactions to
form three flavor quark matter. The advantage is that this does not
require the improbable simultaneous and spontaneous conversion to
strange matter (via weak interactions) of several neutrons within a
small volume. In addition, the Coulomb barrier-free absorption of
neutrons can enlarge the quark phase, so that the whole neutron star
is essentially converted to strange quark matter, thus forming a quark
star. The dynamics of this process is only beginning to be
explored~\cite{Drago}, and it is likely that some mass ejection from
the surface of a neutron star can take place, aided by
neutrinos~\cite{JRK} but powered mostly by pre-compression shock waves
from the rapidly advancing quark conversion front~\cite{JOOM}.

\vskip 0.2cm

There are various ways in which this conversion can be triggered. A
rapidly rotating neutron star is slowed down principally by magnetic
braking (and additionally by energy loss through gravitational waves),
thereby reducing the centrifugal forces and increasing central
pressures. The timescales and dynamics of this spin-down have been
discussed in~\cite{Staff}. The probability of triggering a first order
transition by forming a tiny lump of two flavor quark matter (quantum
nucleation) is exponentially sensitive to the central
pressure~\cite{Bombaci}. This makes the transition much more likely as
the star spins down. Material accumulated from a fallback disk around
a cooling neutron star or accretion from a companion star, for
example, in a low-mass X-ray binary system can also force the
conversion. Even the Bondi accretion rate of $10^{21}$ g/sec from the
interstellar medium can lead to a mass increase of up to 0.1$M_{\odot}$
within a million years which can be sufficient to compress the star
beyond the minimum density for the phase transition. Other
possibilities include the clustering of $\Lambda$-baryons at high
density and seeding by energetic cosmic neutrinos or
strangelets~\cite{AFO}. Deconfinement of quarks can also occur as
early as the protoneutron star stage following a core-collapse
supernova explosion if the central density is large enough. However,
in this case, neutrino trapping in the hot and dense interior can push
the transition density higher, delaying the collapse to the point
where a black hole might be formed instead~\cite{Prakash}.

\vskip 0.2cm

\begin{figure}[ht!]
\centerline{\includegraphics[width=0.7\textwidth,angle=0]{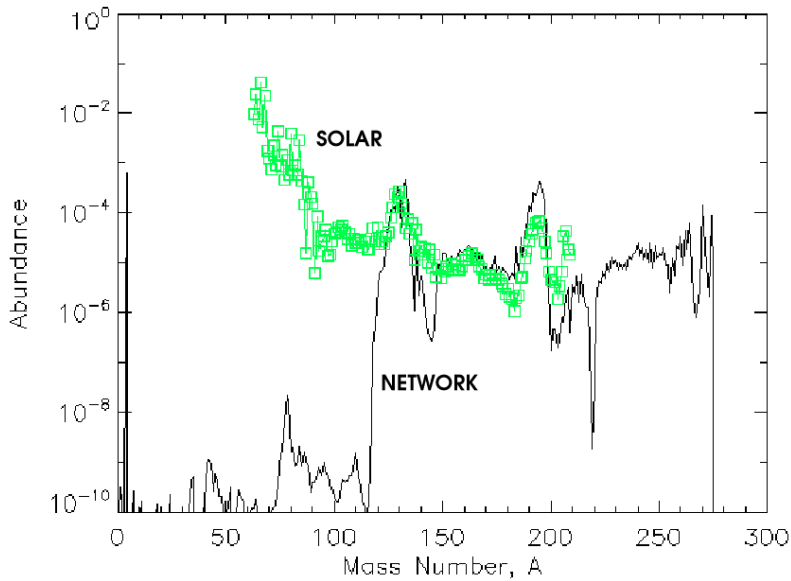}}
\caption{Abundance pattern of $r$-process elements from a Quark-Nova: The
initial electron fraction $Y_e=0.03$, entropy per nucleon $S/k_B=1$ ,
temperature $T_9=1.0$ and expansion timescale $\tau=1.0$
millisecond. The scaled comparison to solar abundance is also shown.}
\label{figps3}
\end{figure}

The amount of mass ejected and the dynamics of the Quark-nova make it
a potential candidate for $r$-process nucleosynthesis. The
decompressing neutron matter at the surface of the neutron star which
is undergoing the phase conversion to a quark star will expand rapidly
and chunks of neutron matter will fall apart, allowing beta-decays to
heat the material to close to $r$-process like temperatures of $T\sim
0.1$ MeV. Following the expansion of this dense ejecta (including
$\beta$-decays) down to neutron drip density using network
calculations of Meyer~\cite{Meyer}, we find that the matter ends up
distributed over a wide range of elements from $Z\sim 40-70$. At this
stage, the full $r$-process reaction network can be coupled to obtain
the final abundance distribution, as well as the evolution of
temperature, entropy and electron fraction. The most important
parameter that governs the evolution of the above physical quantities
is the expansion timescale of the ejecta, which is the time taken to
evolve from a density of $\rho\simeq 10^{14}$g/cc to $\rho\simeq
10^{11}$g/cc. Assuming that the ejecta is on an escape trajectory and
expands as a multiple of the free-fall timescale in a spherically
symmetric, non-interacting manner, we obtain a timescale on the order
of 0.01-1 milliseconds. Fig.~\ref{figps3} shows the $r$-abundance of
the elements (labeled ``network'') from a Quark-Nova compared to the
normalized solar $r$-abundance pattern. The overabundance for elements
with mass number $A>230$ is due to the omission of alpha-decays in the
network. It is also notable that not all neutrons are consumed in this
scenario, and that several light elements are produced with
significant abundance. It is likely that Quark-Novae may be an
additional contributor to the $r$-process material in the universe,
presenting another link between dense quark matter and astrophysics. A
detailed analysis of $r$-process nucleosynthesis from Quark-Novae,
including contrasts to standard or primary astrophysical sites such as
Type II supernovae and neutron star mergers, and comparisons to trends
of heavy-element abundance ratios in metal-poor stars will be
presented in a forthcoming article~\cite{JOOM}.

\vskip 0.2cm
\vskip 0.2cm

While several observable characteristics of strange stars have been
conjectured in the literature~\cite{Glenny,Weber}, such as period-clustering,
anomalies in braking index, global properties (mass and radius), and
small spin-periods, spectral differences between quark stars and
neutron stars have only been touched upon. If strange stars are made
in Quark-Novae, and are linked to $r$-processing of the heavy
elements, one can expect a most direct evidence of a quark star to
come from $\gamma$-decays of unstable nuclei produced in the
$r$-process. We expect the Quark-Nova ejecta to achieve $\gamma$-ray
transparency sooner than supernova ejecta since Quark star progenitors
(i.e., neutron stars) lack extended atmospheres, so that $r$-process
only gamma emitters with lifetimes of the order of years (such as
$^{137}$Cs, $^{144}$Ce, $^{155}$Eu and $^{194}$Os) can be used as tags
for the Quark-Nova.

\vskip 0.2cm

In conclusion, this is an exciting time for quark star
research. Recent developments in the theory of high density quark
matter have provided novel ways of explaining the most energetic
astrophysical events, and spurred re-examinations of old paradigms of
strange stars. With increasing data now coming from myriad sources
(egs. neutron stars that push theoretical limits on spin periods
and radius; GRB spectra that link to quark stars as inner engines;
abundance trends of heavy elements in ultra metal-poor stars that
point to multiple sources of the $r$-process), it appears that quarks
may play an important role in uncovering the secrets of many
astrophysical puzzles.

%-------------------------END TEXT----------------------------%
\section*{Acknowledgments}
I thank the organizers of the IXth workshop on high-energy physics and
phenomenology (WHEPP-9) at the Institute of Physics, Bhubaneswar,
India, for hosting an exciting meeting in a most hospitable setting. I
acknowledge useful interactions with Kaya Mori, Rachid Ouyed, Kaori
Otsuki, Brad Meyer, Mandar Bhagwat and Stewart Wright, and the
hospitality of the Canadian Institute for Theoretical Astrophysics
(CITA), where discussions pertinent to this work were held. P.J. is
supported by the Department of Energy, Office of Nuclear Physics,
contract no. W-31-109-ENG-38. \rm
%-----------------------THE REFERENCES------------------------------------
\begin{flushleft}

\end{flushleft}
\end{document}